\documentstyle[twocolumn,prl,aps]{revtex}
\input epsf     
\begin{document}
\twocolumn[\hsize\textwidth\columnwidth\hsize\csname
@twocolumnfalse\endcsname
\draft
\title{Path Crossing Exponents
and the External Perimeter
in 2D Percolation}
 \bibliographystyle{mabib} 

\author{Michael Aizenman$^a$, Bertrand Duplantier$^b$
and Amnon Aharony$^c$}
\address{$^a$ Departments of Physics and Mathematics, Jadwin Hall,
Princeton University, Princeton, NJ 08544}
\address{$^b$ Service de Physique Th\'eorique de Saclay,
91191 Gif-sur-Yvette Cedex, France   \\
 and Institut Henri Poincar\'e, 11, rue Pierre et Marie Curie 75231
 Paris Cedex 05}
\address{$^c$ School of Physics and Astronomy,
Raymond and Beverly Sackler Faculty of Exact Sciences\\
Tel Aviv University, Tel Aviv 69978, Israel}
\date{Jan. 4, 1999; Revised May 23, 1999.}
\maketitle
\begin{abstract}
$2$D Percolation path 
exponents $x^{\cal P}_{\ell}$ describe
probabilities for traversals of annuli  by  $\ell$ non-overlapping paths,
each on either occupied or vacant clusters, with at least one of each type.
We relate the probabilities rigorously to amplitudes of $O(N=1)$
models whose 
exponents, believed to be exact, 
yield $x^{\cal P}_{\ell}=({\ell}^2-1)/12$.  
This  extends to half-integers  the
Saleur--Duplantier exponents for $k=\ell/2$
clusters,
yields the exact fractal dimension of the 
external cluster perimeter, $D_{EP}=2-x^{\cal P}_3=4/3$, and 
also  explains
the absence of narrow gate fjords, as originally found
by Grossman and Aharony.

\end{abstract}
    ]

The fractal geometry of critical percolation clusters has  been of
interest both for intrinsic reasons and as a window
on a range of phenomema.  It is characterized by fractal
dimensions of various sets \cite{book,stanley}, e.g.,
of the connected clusters,  their backbones,
the  sets of pivotal (singly--connecting)
bonds,  the clusters' boundaries (hulls),
and their external (accessible) perimeters.
A set  $S$ is said here to be of fractal dimension
$D_{S}$
if the density of
points in $S$ within a box of linear size $R$ decays as
$R^{-x_{S}}$, with $x_{S}=d-D_{S}$ in $d$ dimensions.
 
For two--dimensional (2D) independent percolation, many of the
fractal dimensions have been found exactly
\cite{Denijs,N,SD,Cardy}, though most of these values
have not yet been established at a rigorous level.
In several cases, Saleur and Duplantier (SD)\cite{SD} identified
the co--dimension $x_{S}$
with the exponent $x^{\cal C}_{k}$ which describes the decay 
law $P^{\cal C}_{k} \approx (r/R)^{x^{\cal C}_{k}}$ 
for the probability ($P^{\cal C}_{k}$)
%
that in an annular region $D(r,R)$ 
the small
circle of radius $r$ is connected to the outer one, 
of radius  $R >> r $, 
by $k$ different clusters of occupied sites (or bonds).
SD utilized the observation that the statistics of the
$2k$ {\it boundary} lines of the connected clusters
correspond to those of loops in some  well recognized models:
the $Q=1$  Potts model (at its critical point)
for the bond percolation model and the $O(N=1)$
loop model of Domany {\it et al.}\cite{Domany}
 (at its {\it low temperature
phase}) for site percolation on the triangular lattice.
Using the ``Coulomb gas'' representation
for the corresponding $\ell$-line exponents,  $x^{O(N)}_{\ell}$,
SD obtained \cite{SD} for both models the
values, expected to be universal, 
\begin{equation}
x^{\cal C}_k = x^{O(N=1)}_{\ell=2k} = (4k^2-1)/12 \; ,
\label{eq:DS}
\end{equation}
where $k$
{\it clusters} correspond to $\ell=2k$
{\it lines} in the loop model.  

Among the noteworthy applications of the above formula are
the ``hull dimension'', i.e., the dimension of the
cluster's perimeter,
\begin{equation}
D_{H}=2-x^{\cal C}_{1}=7/4,
\end{equation}
and the dimension of the set of ``red'' (singly connecting) bonds,
which are pinching points between two large clusters:
\begin{equation}
D_{SC}=2-x^{\cal C}_2=3/4=1/{\nu},
\end{equation}
where $\nu$ is the correlation length exponent
\cite{coniglio}, in agreement with
previously derived values \cite{Denijs}.
However, some well known percolation dimensions 
have eluded 
this exact approach:
the dimension $D_{EP}$ of the {\it external (accessible) perimeter} (EP)
or frontier of a
cluster, first studied by Grossman and Aharony (GA)\cite{GA},
and the {\it backbone} dimension. The EP of a cluster
is the {\it accessible} part of the hull, which excludes
deep ``fjords'' which are connected to the cluster's complement
through very narrow passages
(or ``gates"). The dimension of the EP
was found numerically
to be $D_{EP}\approx 4/3$ \cite{DEP}.
  GA \cite{GA} also made the puzzling observation  that,
while typical clusters do show many fjords with only a narrow passage
to the complement,
once one fills in fjords with passages of width
two or three lattice spacings -- no fjords  of broader microscopic
passages, and depth comparable with that of the cluster, are left.
This is clearly visible in Fig. 6 of the second Ref. \cite{GA}.
Both of these observations make the EP look very similar to self--avoiding
walks
(SAW's).  Although there appeared conjectures attempting to make this
relation quantitative \cite{SD,GA}, the
connection was never elucidated.

  In this Letter we report on a resolution of these issues
  through analysis of the {\it path crossing} probabilities.

\noindent {\it i. \/}
Basing the relation of percolation exponents with the
$O(N=1)$  exponents on a somewhat different footing than that used in
Ref.\cite{SD}, we  extend the list of exact values proposed for critical
percolation in 2D.
Instead of focusing on entire clusters, we consider the probability
$P^{\cal P}_\ell(r,R;\tau_1,\ldots, \tau_{\ell})$ that
the annulus $D(r,R)$
is traversed by (at least) $\ell$ non--overlapping connected
{\em paths},  which are ``monochromatic'' in the sense that each
consists of either occupied sites (``color'' $\tau_j = +$) or
vacancies ($\tau_j = -$).
We rigorously prove~\cite{JMP} that for color sequences which include
at least one of each type ($\pm$) the decay rates of the probabilities are
{\it color-independent},
and are given by the $O(N=1)$ exponents.
Assuming the validity of the exact
values for the latter,  we find that
\begin{equation}
P^{\cal P}_\ell(r,R;\tau_1,\ldots, \tau_{\ell})
\approx (r/R)^{x^{\cal P}_{\ell}}
\end{equation}
with the {\it path crossing exponents} $x^{\cal P}_{\ell}$ satisfying
\begin{equation}
x^{\cal P}_\ell = x^{O(N=1)}_{\ell} = (\ell^2-1)/12.
\label{eq:path}
\end{equation}
Since  the cluster exponents are $x^{\cal C}_{k} =  x^{\cal P}_{2k} $,
Eq.~(\ref{eq:path}) may be viewed as
an extension of the SD formula to {\it odd} values of $\ell$, or
{\it half integer} values of $k$, and to more general
sequences $\tau_j=\pm$.

\begin{figure} 
      \begin{center}
       \leavevmode
    \epsfxsize=2.5in
      \epsfbox{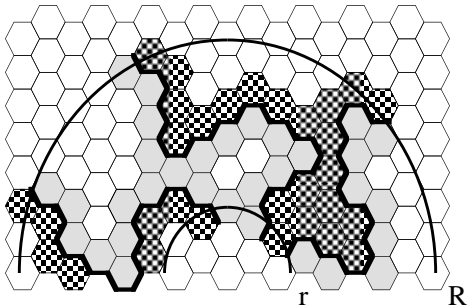}
  \caption{
  A configuration with $\ell = 2$  paths  traversing
  the semi-annular region $\tilde{D}(r,R)$, one of occupied
  (checkered) and the other (dual) of vacant (gray) hexagons.
  The configuration has
  $(\ell+1 = 3)$  $\ O(N=1)$ boundary lines (drawn thickly);
  the states of the  hexagons
 not adjacent to these lines  are left unspecified.}
  \end{center}
  \end{figure}

\medskip
\noindent {\it ii. \/}
Using the newly
acquired values we explain
some of the quantitative and qualitative features of the EP of
critical clusters mentioned above: its dimension, which we
identify as
\begin{equation}
D_{EP}=2-x^{\small O(1)}_{3}= 4/3 \; ,
\label{dimEP}
\end{equation}
and the interesting fact that -- unlike the hull -- the EP
appears to be self--avoiding on the macroscopic scale.

\medskip

\noindent {\it iii. \/}
We consider also the analogous
{\it boundary} or {\it ``surface'' exponents\/}, which describe
the probability
$\tilde{P}^{\cal P}_\ell(r,R;\tau_1,\ldots, \tau_{\ell})$
that, {\it within the upper half space},
a semi-annular region $\tilde{D}(r,R)$
is traversed by  $\ell $ paths (see Fig. 1).
For the exponents defined by
\begin{equation}
\tilde{P}^{\cal P}_\ell(r,R;\tau_1,\ldots, \tau_{\ell})
\approx (r/R)^{\tilde{x}^{\cal P}_{k}}
\label{eq:P-surface}
\end{equation}
we find
\begin{equation}
\tilde{x}^{\cal P}_\ell = \tilde{x}^{O(N=1)}_{\ell+1} =
(\ell+1) \ell/6.
\label{eq:surface}
\end{equation}
In this case the relation
is valid with {\it no} restriction
on the color sequence $\tau$;  however there is a shift: $\ell$
crossing paths correspond to $(\ell+1)$
$O(N=1)$ lines. Thus, with odd $\ell$, one recovers the cluster
boundary exponents $\tilde{x}^{\cal C}_{k} = \tilde{x}^{\cal
P}_{2k-1}=k(2k-1)/3$, as in Refs.~\cite{BD,Cardy98}.

Before we turn to describe the arguments for the exact values of the path
exponents, as provided by Eqs. (\ref{eq:path}) and  (\ref{eq:surface}),
let us  present their implications
concerning the dimension and shape of the external perimeter.
Each point on the accessible EP
is next to the end of three paths
of   lengths comparable with the diameter of the cluster -- a path
of occupied sites and in addition two distinct dual paths
of vacancies (Fig. 2a),
which guarantee that the point is not within a fjord of narrow opening
(both paths must be able to exit the fjord via the narrow gate).
This yields
$D_{EP} \ = \  2 - x^{\cal P}_3 = 4/3,$
in excellent agreement with the numerical results\cite{DEP}.

\begin{figure}
    \begin{center}
     \leavevmode
    \epsfysize=1.4in
    \epsfbox{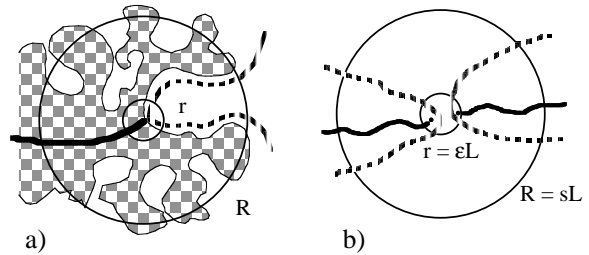}  
  \caption{ Paths, and dual (dotted) paths  characterizing
  a) the external perimeter  ($\ell =3$) and b) a gate
  ($\ell =6$).}
  \end{center}
  \end{figure}

GA's observation
concerning the fjords is particularly
striking from the perspective of the {\it scaling limit}, for which one
sends the lattice spacing to zero while keeping the sight on the
curves observed on the macroscopic scale.
(The limit can be constructed using the analysis of
Ref.~\onlinecite{AB},
which implies that
 the cluster hulls and EP's can still be described
 by means of H\"older continuous random curves.)
While the EP is self--avoiding on
the lattice scale, like  the hull  it could have
close encounters which appear as self--intersections when
viewed from the macroscopic perspective.  Yet such close encounters
are not observed.
Also this puzzle is explained by the generalized path statistics:
the occurence in an $L\times L$ box  of a
cluster with a fjord of depth $R=s L$  and neck
width $r=\epsilon L$ requires there being {\em six} paths,
two pairs of triplets as used in the derivation of Eq. (\ref{dimEP}),
which meet in a region of size
$r$ and avoid each other up to a radius $R$ (Fig. 2b).
The probability of finding such six paths scales as
$  \epsilon^{-d} \times
 \left({\epsilon \over s }  \right)^{x^{\cal P}_6}  \ = \
 O(\epsilon^{x^{P}_6 -d})$
where $d=2$.
Equation (\ref{eq:path}) yields the exponent
value
$x^{\cal P}_6=2 {11 \over 12}$
and hence the probability for a randomly picked configuration to
exhibit such a gate   tends
to zero, in the situation where $s$ is
fixed at some non-infinitesimal value  $ 0 < s < 1$, and
$\epsilon \to 0$.
The crucial point here is that
the fractal dimension of the set of these gates  is negative:
\begin{equation}
 D_G\ =\ 2-x^{\cal P}_6 \  < \  0 \; .
\label{DG}
\end{equation}
This explains the asymptotic absence of macroscopic
size fjords of neck width $r$
anywhere between few multiples of the lattice spacing
and $\epsilon L$, for any fixed $ \epsilon \ll 1$.
The above argument also implies that the EP will
not exhibit {\it peninsulas} with narrow isthmuses, even though that
condition was not built into the construction.

It is of interest to recall here the suggestion which was made on
a theoretical as  well as a numerical basis \cite{CJMS,DS1}, that
percolation's hull and EP dimensions coincide with
the dimensions of polymers, respectively at the $\theta$-point (the
onset of collapse),
or in  the SAW  state:
\begin{equation}
 D_{H}=D_{\theta}, \quad D_{EP}=D_{SAW} \; .
\end{equation}
It is natural to conjecture that in the scaling limit the EP
{\it coincides}, in its local statistics,
with a SAW.
This may appear to  be in
conflict with the a-symmetry
between the two sides of the EP, however
the absence of peninsulas in the scaling limit suggests that
the symmetry 
is  restored asymptotically.

The above results (\ref{eq:path}-\ref{DG}) are based on a rigorous
relation of path crossing
probabilities with amplitudes
of a loop model which we shall now define, and on known exact values
for the exponents of the latter (which still remain
to be proven at a rigorous level).
The arguments, which will be presented more completely in
ref.~\cite{JMP}, are formulated for the special model of independent
site (i.e., hexagon) percolation on the triangular lattice.
For reasons of universality one may expect the conclusions
to apply to other 2D percolation models,
e.g., the bond percolation,  and also to the
statistics of the connected clusters of ($+$) or ($-$) spins in
the 2D Ising model at all temperatures above $T_c$.

The loop-model configurations, $\Gamma$,  are collections of
nonoverlapping loops and lines, in suitable subsets of the
plane,  which are allowed to have
end-points  only  within prescribed regions.
The weight of a configuration is
\begin{equation}
W(\Gamma) \ = \ K^{\cal N_B} N^{\cal N_P}  \;,
\end{equation}
with  ${\cal N_P}$ the number of closed lines (or ``polygons'')
and ${\cal N_B}$ their total length (the number of bonds).
For the particular case of hexagon percolation discussed here
the fugacities are $K=1$,
$N=1$, i.e.,
$W(\Gamma) = 1 $ for all $\Gamma$ \cite{SD}.

A probability distribution
of loop and line configurations  in a prescribed region
is defined by means of the weights
$W(\Gamma)/Z$, with $Z$ a suitable normalizing  factor.
Let now  $P^{O(N)}_{\ell}(r,R)$ denote the
probability that such a system of lines with no end points
in the annular domain $D(r,R)$ contains at least $\ell$ lines
traversing $D(r,R)$.
For a representation of the surface exponents,
we also let $\tilde{P}^{O(N)}_{\ell}(r,R)$ denote the corresponding
event with the lines restricted to lie in the upper half
plane, taken here  with the ``free boundary conditions''.
  A close variant of the quantity $P^{O(N)}_{\ell}(r,R)$  is the
$O(N)$ {\it amplitude}
$G^{O(N)}_{\ell}(r,R)$ which is defined as the
 {\it sum} over $\ell$ lines
$\{\gamma_1,\ldots, \gamma_{\ell}\}$ spanning the annulus
$D(r,R)$ of the {\it probability} $\omega(\gamma_1,\ldots,
\gamma_{\ell} \in \Gamma)$
 that the lines are included
in $\Gamma$.  For $N=1$, that probability
reduces to the {\it local} expression\cite{DS1,JMP}:
\begin{eqnarray}
\omega(\gamma_1,\ldots, \gamma_{\ell})
&=&2^{-{\cal N}_{H}(\gamma_1,\ldots, \gamma_{\ell})}
2^{{\cal K}(\gamma_1,\ldots, \gamma_{\ell})} / 2^{\#}
\end{eqnarray}
with ${\cal N}^H(\gamma_1,\ldots, \gamma_{\ell})$ the number
of hexagons touched by the lines,
${\cal K}(\gamma_1,\ldots, \gamma_{\ell})$ the number of
line clusters -- two lines being regarded as
 in the same cluster if they touch a common hexagon,
 and  $\#$ defined as taking the value $0$ if the $\ell$ lines leave
room for another curve to
traverse  the annulus and  $1$ otherwise. The amplitudes then read
\begin{equation}
G^{O(N=1)}_{\ell}(r,R) \  = \ \sum_{ \gamma_1,\ldots, \gamma_{\ell} }
\omega(\gamma_1,\ldots, \gamma_{\ell}),
\end{equation}
the sum running over sets of $\ell$ nonoverlapping lines
which traverse $D(r,R)$.
It can be shown that the probabilities and amplitudes
agree to the leading order \cite{JMP}:
\begin{equation}
P^{O(N=1)}_{\ell}(r,R) \  = \ \left(1 + o({r \over R})
\right)G^{O(N=1)}_{\ell}(r,R) \;.
\end{equation}

``Coulomb gas'' and Bethe Ansatz methods \cite{N,SD,BB,DS,BS93} yield
the
conclusion that the loop model amplitudes, and 
thus also the probabilities, decay
by power laws,
$G^{O(1)}_{\ell}(r,R)  \approx
(r/R)^{x^{O(1)}_\ell}$, and
$\tilde{G}^{O(1)}_{\ell}(r,R)  \approx
(r/R)^{\tilde{x}^{O(1)}_\ell}$,
with the exponents taking the values
given in Eq.~(\ref{eq:path})
and Eq.~(\ref{eq:surface}).
Our results rest now on the fact that the
$O(N=1)$ line  probabilities are of the same order
of magnitude as the path crossing probabilities.
Their compatibility is expressed in the following statement.

\medskip

\noindent
{\bf Proposition} \/ {\em  In the site percolation model on the
triangular lattice:

1)  For any ``color sequence''
$\{\tau_j=\pm\}_{j=1}^{\ell}$ which
includes at least one of each kind ($+$ and $-$),
\begin{equation}
P^{\cal P}_\ell(r,R;\tau_1,\ldots, \tau_{\ell})  \  \
{\stackrel{\scriptstyle <}{_{_{\scriptstyle >} }} } 
 \  \  P^{O(N=1)}_{\ell}(r,R)  \;  ,
\label{eq:general}
\end{equation}
where $A \ {\stackrel{\scriptstyle <}{_{_{\scriptstyle >} }} } \ B$
means that there are constants $0 < c_1, c_2 <
\infty$ with which $c_1 A \le  B \le c_2 A$ {\it uniformly in $r$ and
$R$}.

2)  The surface probabilities satisfy
\begin{equation}
\tilde{P}^{\cal P}_\ell(r,R;\tau_1,\ldots, \tau_{\ell}) \ = \
\tilde{P}^{O(N=1)}_{\ell+1}(r,R)  \;  ,
\label{eq:Gtilde}
\end{equation}
without any restriction on the color sequence $\tau$.
 }
\medskip

Let us outline here the proof, whose details will be spelled in
ref.~\cite{JMP}.
The simplest case of the above relation is in the example of
the half-disk amplitude with alternating color paths (as in Fig. 1),
which corresponds to
$\tilde{P}^{\cal P}_\ell(r,R;+,-,+,-,\ldots)$.
Equation (\ref{eq:Gtilde}) holds there since the statistics
of the boundary lines is given exactly by the $O(N=1)$ loop model.
The result is then extended by
establishing independence on the color sequence.
This is done by successively conditioning on the suitable
``rightmost path'' and flipping the site variables left of the
line.  Thus use is made of the Markov property combined with
 the spin flip symmetry, which are enabled by the independence
 and the self duality of the site percolation model on the
 triangular lattice.
The argument is a bit more involved in
the case of the full disk.
There we need to have at least one traversing boundary line, which
 we employ to slit the annulus.  The previous argument is then
applied to the resulting simply connected domain.
Equation ~(\ref{eq:general}) reflects
the fact that the overcounting involved in the selection of
the slit is by at most a finite factor.

As noted in \cite{Kesten_scaling,Aiz_Xiamen}, it is possible to
obtain some
selected path exponents by direct arguments.
The values agree with the formulas given above.
It is instructive to list  specific values of
  $x^{\cal P}_{\ell} = (\ell^2 - 1)/12$: \\
%
%
%
\mbox{   }$ \ell =2$:  \quad  $x^{\cal P}_{2}$ yields
$D_{H}=D_{\theta}=7/4$.
\\
\mbox{   }$ \ell =3$:  \quad   $x^{\cal P}_{3}$ yields
$D_{EP}=D_{SAW}=4/3 $. \\
\mbox{   }$ \ell =4$:   \quad   $x^{\cal P}_{4}$ yields
$D_{SC}=\nu^{-1} =
3/4$.
\\
\mbox{   }$ \ell =5$:  \quad    $x^{\cal P}_{5}=2$ can be derived
directly.  \\
\mbox{   }$ \ell =6$:  \quad    $x^{\cal P}_{6} > 2 $  implies that
the EP is self--avoiding on the large scale ($D_G <
0$).

The relation (\ref{eq:path}) was not claimed for $\ell=1$, or for
paths of a single color ($\hat{x}^{\cal P}_{\ell}$).
Concerning this let us note: \\
$\ \  $ -- It can be shown directly that $x^{O(N=1)}_{1}=0$
\cite{JMP} while
 $\hat{x}^{\cal P}_1 > 0$ \cite{pi}.
 The path exponent is related to the cluster dimension; its
value appears to be $\hat{x}^{\cal P}_{1} = 5/48 $ \cite{Denijs}. \\
$\ \ $ -- The case of $\ell = 2$,
 with two paths of the same color,
is of
special interest since it relates to the {\it backbone dimension}.
Numerically,
$\hat{x}^{\cal P}_{2} = 0.3568\pm 0.0008$ \cite{grass}.

For the surface exponents,
$\tilde{x}^{\cal P}_\ell = (\ell+1) \ell/6$  of
Eq.~(\ref{eq:surface}), we note
that  for $\ell \to \infty$ \
$\tilde{x}^{\cal P}_\ell / x^{\cal P}_\ell \to 2 $,
as it should; and
\mbox{   }$ \ell =1$:  \quad   $\tilde{x}^{\cal P}_1 = 1/3 \ \  $
is consistent
with Cardy's equation for the crossing probability \cite{Cardy}.
  \\
\mbox{   }$ \ell =2$:  \quad    $\tilde{x}^{\cal P}_2 = 1$  can be
derived directly. \\
\mbox{   }$ \ell =3$:  \quad    $\tilde{x}^{\cal P}_3 = 2$  is also
directly derivable. \\
The last one  \cite{Aiz_Xiamen} is related to
a slit-disc exponent which is attributed to   
J. van den Berg in Ref. \cite{Kesten_scaling}.




Finally, we note that the SD formalism also yields predictions
for the hull dimensions of Fortuin-Kasteleyn random clusters,
describing
the $Q$-state Potts model.  These  were recently confirmed
in numerical simulations by Hovi and Mandelbrot \cite{HM}.
In contrast, the  values found in that work for the external
perimeters  do not agree with the generalizations
of the SD formulas to odd $\ell$.   The
results presented  here were derived only for site percolation.
It would be interesting to see generalizations.

%
This paper is dedicated to the memory of Tal Grossman.
The work was started and carried out while the authors enjoyed the
gracious hospitality of the Institut Henri Poincar\'e,  the Institute
for Advanced Studies (MA and BD), and of Tel Aviv University (MA).
It was supported in part by the NSF Grant PHY-9512729 (MA),
a grant from the German Israeli Foundation (AA), and
by a grant to the IAS from the NEC Research Institute.


\end{document}